\documentclass[fleqn,twoside]{article}
\usepackage{espcrc2}
\usepackage{epsfig}
\newcommand{\ave}[1]{\left\langle #1 \right\rangle}
\newcommand{\tilro}{\tilde{\rho}}
\newcommand{\order}[1]{ \mathcal{O} \left( #1 \right) }

\usepackage{graphicx}
\usepackage[figuresright]{rotating}

\newcommand{\AmS}{{\protect\the\textfont2
  A\kern-.1667em\lower.5ex\hbox{M}\kern-.125emS}}

\title{Heavy quarks and the collective properties of hot QCD}
\author{G.Torrieri\address[FIAS]{FIAS, Johann Wolfgang Goethe
Universit\"{a}t,Frankfurt A.M., Germany}%
        \thanks{Work was financially supported by the Helmholtz
International
Center for FAIR within the framework of the LOEWE program
(Landesoffensive zur Entwicklung Wissenschaftlich-Ökonomischer
Exzellenz) launched by the State of Hesse.},
J.Noronha\address[CU]{Department of
Physics, Columbia University, 538 West 120$^{th}$ Street, New York,
NY 10027, USA}%
        \thanks{Acknowledges support from DOE under
Grant No.\ DE-FG02-93ER40764.  }}
\begin{document}

\begin{abstract}
After reviewing the evidence that QCD matter at ultrarelativistic energies behaves as a very good fluid, we describe the connection of QCD fluidity to heavy quark observables.   We review the way in which heavy quark spectra can place tighter  limits on the viscosity of QCD matter.   Finally, we show that correlations between flow observables and the event-by-event charm quark abundance (``flavoring'') can shed  light on the system's equation of state \cite{ourpaper}.
\vspace{1pc}
\end{abstract}

\maketitle

The purpose of high energy heavy ion collisions is to study the thermodynamic properties of strongly interacting matter at high temperatures and densities.  The most ambitious part of this program is to create a bubble of deconfined ``quark-gluon plasma'', a gas of ``free'' quarks and gluons mimicking the properties of the universe shortly after the big bang \cite{jansbook}.

This program, however, has a potential obstacle:  Thermodynamics generally applies to ``large'',''static'' systems, where ``large''  is compared to   typical sizes of microscopic degrees of freedom.    While the system created in a heavy ion is large compared to the quark or hadron size, the system is certainly not static.  In fact, it {\em could} be be a far-from-equilibrium ``mess of quarks and gluons'', where each parton scatters a few times, but whose dynamics has no connection to thermodynamics.  In this case, concepts like phase transition have little value.

The best that we can hope for is to replace ``static'' by ``slowly evolving'', where ``slowly'' is defined in respect to the timescales of the microscopic processes within the system.   If the system is slowly evolving (alternatively, if microscopic dynamics equilibrates very fast, as it naively should in a strongly coupled hot system), we can explore the thermodynamics of the system via hydrodynamics:The systems collective motion will be governed by the equation of state and transport coefficients, calculated from equilibrium thermodynamics.

RHIC experimental data has given us reason for optimism in this respect:
One of the most widely cited findings in ultrarelativistic heavy ion 
collisions concerns the discovery of a ``perfect fluid'' in collisions
at the Relativistic Heavy Ion Collider (RHIC)
\cite{v2popular,sqgpmiklos}. The
evidence for these claims comes from the successful modeling of the
anisotropic expansion of the matter in the early stage of the reaction
by means of ideal
 hydrodynamics \cite{heinz,shuryak,huovinen}

It appears, therefore, that the system is, to a very good approximation, locally thermalized, and we can use flow observables as a probe of the equation of state,and eventually of phase transitions.    However, efforts in this direction have met remarkably little success:   The probes for which hydrodynamics works very well are insensitive to the equation of state \cite{shuryak,huovinen}.  Probes (such as HBT radii and average transverse momentum) which are more sensitive show no structure indicative of a phase transition \cite{heinz,shuryak,huovinen},presumably because the systems initial conditions are not ``tidy'' enough to highlight discontinuities in the equation of state \cite{pratt,larry}.

Other probes more sensitive to the details of the equation of state are therefore needed.    One obvious candidate is heavy quark observables.
A heavy probe in a thermalized medium is, as is well-known, described by Brownian motion.  This makes it a very sensitive probe of thermalization, since the scale of thermalization of a heavy particle $\tau_{heavy}$ is parametrically larger wrt to its light component $\tau_{light} \sim \eta/(Ts)$ (where $\eta/s$ is the dimensionless ratio of viscosity to entropy density and $T$ is the temperature): If the mass of the heavy particle is $M\gg T$, $\tau_{heavy}/\tau_{light} \sim M/T$ \cite{teaneyc}.

RHIC data to date is consistent with ``heavy particles flowing as much as light particles'' \cite{phecharm,whitestar}, although large systematic errors persist 
(for example, we can not distinguish a charm from a bottom quark)
If this is really true, than $\eta/s$ is truly $\ll 1$ and heavy quarks are, for all intents and purposes, thermalized with respect to the rest of the system.  They (in particular, the more abundant charm quarks) can then be used as chemical probes for the equation of state.
\begin{figure}[h]
\epsfig{width=6cm,clip=,figure=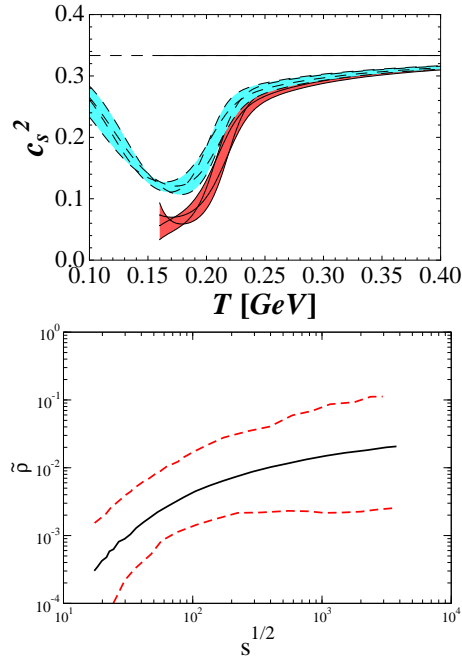}
\epsfig{width=6cm,clip=,figure=fig_rhotilde.eps}
\caption{(color online) \label{saltcs} Top panel: The $c_s^2$ dependence temperature, for 
$\tilro=0$ (dashed lines) and $\tilro=0.1$ (solid lines).  The width of the bands denotes lattice
uncertainties \cite{bazavov} in the speed of sound of a 2+1 QGP, while the thick line denotes the conformal,
non-interacting value where $c_s^2=1/3$.
Bottom panel: \label{tilderho} $\tilro$ vs $\sqrt{s}$ for Pb-Pb collisions computed using Eq.\ 
(\ref{tilderho}).  The dashed lines denote the uncertainties in the perturbative QCD calculation   }

\end{figure}

In general, charm in heavy ion collisions is not expected to be {\em chemically} equilibrated.   The bulk of charm content should be produced by ``hard'' processes in the initial state at a concentration far above their equilibrium expectation \cite{ramona}.   The abundance of $c \overline{c}$ pairs produced in heavy ion collisions at the LHC is expected to be reasonably high \cite{ramona} ($\sim 10^{1-2}$, of course parametrically smaller than the total $\sim 10^{4}$ multiplicity).    
In the dilute limit, the total charm abundance is for all intents and purposes a conserved number.  Hence, a good observable is the dimensionless quantity $\tilro = \rho/s \sim \rho T^{-3} \ll 1$.  The events entropy, also nearly conserved for a good fluid, can be related to the multiplicity rapidity density $dN/dy$ \cite{jansbook}.

In the dilute (corrections of $\order{\tilro^2}$) infinitely heavy quark (corrections of $\order{T/M}$) limit, the contribution of an abundance of charm quarks to the free energy can be computed by adding a Polyakov loop density \cite{polyakov} to the free energy density
\begin{equation}
\mathcal{F}(T) = \mathcal{F}_0(T) +\tilro \,s_0(T) \, F_Q (T)
\label{noronhaeq1}
\end{equation} 
where $F_Q(T) = -T \ln \ell(T)$ and $\ell(T)$ is the renormalized Polyakov loop.
The Polyakov loop is obtainable from lattice calculations, and the quantities underscored with 0 denote the values before the charm flavor was included.  The speed of sound can then be computed by textbook thermodynamic methods,
$c_s^2 = \frac{d\ln T}{d\ln s}\,$, and $s=-dF/dT$.
 Fig. \ref{saltcs} top panel shows our estimate for the speed of sound derived via Eq.\ (\ref{noronhaeq1}) using the expectation value of the Polyakov loop extracted from the lattice  (2+1 QGP with almost physical quark masses \cite{bazavov}). One can see that the main effect comes from the region near the phase transition (where there is a minimum in the speed of sound) but well before the Polyakov loop expectation value reaches its asymptotic high-T limit, leading to a negative shift of the speed of sound from its value in a 2+1 QGP. This can be physically readily understood:   Correlations between the medium and slowly moving heavy quarks lower the system's response to pressure.

Our estimate stops close to $T_c$, as $-T \ln \ell(T) \rightarrow \infty$ in the confining phase (where the Polyakov loop expectation value vanishes).    Mathematically, one can trust our approach as long as the heavy quark is much heavier than any other scale in the system, ie $-T \ln \ell(T) \ll M_q$.  At some point in the approach to confinement, however, this approximation breaks down.
To estimate the contribution of flavoring in the confined phase we assume flavorful confined QCD is described by the hadron resonance gas model.    In this case, flavoring can be approximated by an admixture of heavy mesons in a gas of pions.  The latter has a speed of sound of $c_s^{light} \simeq 1/\sqrt{3}$ (ultra-relativistic ideal gas), while the former will have a speed of sound of $c_s^{heavy} \simeq \sqrt{5T/(3M_{meson})}$.  The speed of sound of the mixture will therefore go as
\begin{equation}
\label{csweak}
c_s^2 \sim \frac{1}{3} + \order{\tilro \frac{T}{M_{meson}}}
\end{equation}
parametrically smaller than the contribution in the deconfined phase, which is just $\order{\tilro}$.    Thus, the flavoring effect on the speed of sound is
{\em specific to the deconfined phase}.   The effect of flavoring in a {\em weakly coupled} QGP is similarly described by Eq. \ref{csweak} (with $M_{meson} \rightarrow M_{charm}$), so a correlation of $c_s$ with flavoring would indicate a deconfined but strongly coupled system.

\begin{figure*}[t]
\epsfig{width=12cm,clip=,figure=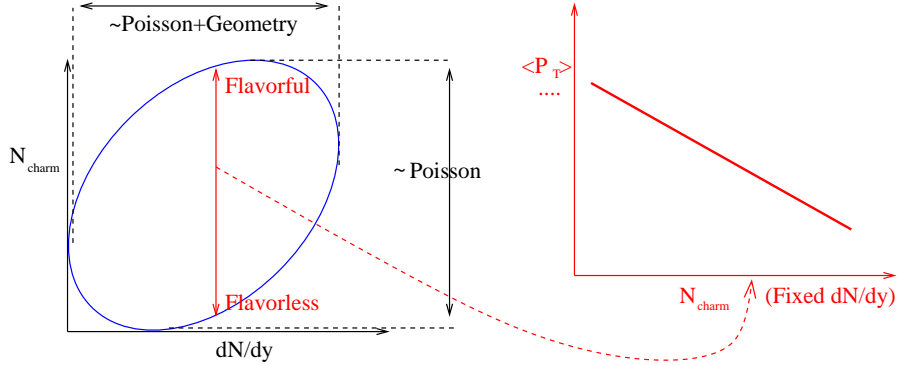}
\caption{(color online) \label{saltthermo} Left panel: The  event-by-event distribution of events with 
respect to charm content and $N_{part}$, together with the cut required to analyze the response of the system to
charm.  Right panel: The  dependence of $\ave{p_T}$ with charm number.  }
\end{figure*}
These effects produce observable consequences.   Using pQCD estimates \cite{ramona} and logarithmic scaling for multiplicity rapidity density, we estimate $\tilro$ at the LHC to be (Fig. \ref{saltcs} bottom panel)
\begin{equation}
\tilro = \frac{1}{6}\frac{dN_{charm}/dy}{dN_{charged}/dy} \simeq
\label{tilderho}
\end{equation}
$$ \frac{1}{3} \frac{N_{collisions}}{N_{participants}} \frac{\sigma_{pp \rightarrow c \overline{c}}(\sqrt{s}) \Lambda_{QCD}^2}{ \Delta y \ln \left( \frac{\sqrt{s}}{E_0} \right)} \sim 0.05$$
This is the average expectation. $dN_{charm}/dy$ and $dN/dy$ will however vary event by event, in an approximately Poissonian manner.  Provided that charm can be reasonably reconstructed and there is a large enough event sample, $\tilro$ is an {\em experimental observable} capable of serving as a binning class for events (see Fig. \ref{saltthermo}).

As is well known, there is a connection between the speed of sound and the limiting average velocity of a hydrodynamic expansion with shock-like initial conditions, 
$\ave{\gamma_T v_T}_{freezeout} \sim f(N_{part}) \ave{c_s}_{\tau}^2$
where ``freezeout'' implies averaging over the freeze-out hypersurface \cite{jansbook} while the subscript $\tau$ means the average is done over the hydrodynamic evolution. 
For a shallow shock this result is exact \cite{dirkhydro1}. 
While knowledge of the initial geometry is needed to establish the form of $f(N_{part})$, model calculations \cite{shuryak} indicate that the dependence is not washed away even in steeper shocks and more complicated initial geometries. 

  The final transverse flow is in return connected to the average transverse momentum
$\ave{p_T} \simeq T + m \ave{\gamma_T v_T}$.
Hence, the decrease of the speed of sound close to $T_c$ (Fig. \ref{saltcs}) could lower $\ave{p_T}$ for more flavored events with respect to flavorless ones (Fig. \ref{saltthermo} right panel). Note that this effect is {\em opposite} to the naturally expected positive correlation due to the correlation between $N_{collisions},N_{part}$ and $dN/dy$.  The coefficient associated with this heavy flavoring effect would be straightforwardly related to non-perturbative QCD via Fig.\ \ref{saltcs}.   This effect might be easier to measure in smaller systems due to the greater event-by-event variation in $\tilro$ and less background.
  The main requirement of such an analysis is the ability to experimentally gauge both the charm quark abundance and $\ave{p_T}$ {\em event-by-event}

A possible ``trivial'' effect which would give correlations in the same sense as the 
effect proposed here is energy conservation (roughly, charm quarks need
a lot of energy to be created, and that lowers $\ave{p_T}$).
 The correlation due to energy conservation should,however, be suppressed by factorization and
boost-invariance: Charm quarks are created of partons with larger
Bjorken $x$ \cite{feybj} than the mid-rapidity soft particles accounting for most of $dN/dy$, so the energy they take up
comes from regions where rapidity deviates from zero.  We do not expect,
therefore, that energy conservation will lower $\ave{p_T}$ at
mid-rapidity.  Hence, the observation of a significant charm-flow correlation should be due to the collective properties of the system, {\em not} energy conservation.

In conclusion, we have argued that heavy quark observables seem to confirm that the matter produced in RHIC collisions is a very good fluid.  This means that heavy quarks are, too a good approximation, in thermal (but not chemical)  equilibrium with the rest of the medium.  This makes it possible to study the medium response to the presence of heavy quark impurities.  This response is calculable, to a good approximation, from lattice QCD data, and can produce non-trivial correlations between event charm abundance and flow properties.

\end{document}